\documentstyle[12pt]{article}
\def\be{\begin{equation}}
\def\ee{\end{equation}}
\def\sd{s^{\dagger}}
\def\dd{d^{\dagger}} 
\def\gd{g^{\dagger}} 
\def\dt{\tilde{d}}
\def\gt{\tilde{g}} 
\begin{document}

\section {Introduction}

In the literature fermion pairs are replaced by bosons in many known
physical situations. This is normally performed 
with the help of boson mappings, that link the fermionic Hilbert
space to another Hilbert space constructed with bosons. Of course
boson mapping techniques are only useful when the Pauli Principle
effects are somehow minimized. Historically boson expansion theories
were introduced from two different points of view. The first one
is the Beliaev - Zelevinsky - Marshalek (BZM) method \cite{bzm}, which 
focuses on the mapping of operators by requiring that the boson images
satisfy the same commutation relations as the fermion operators. 
In principle, all important operators can be constructed from a set of
basic operators whose commutation relations form an algebra. The 
mapping is achieved by preserving this algebra and mapping these basic 
operators. The Dyson mapping \cite{dyson} is of the BZM type. It provides finite
boson expansions but non - hermitian boson images. The boson expansions
of the multipole operators contain only two-boson terms, while the 
expansions of the pair operators contain only one- and three- boson
terms. In spite of being non- hermitian, boson images can be hermitized,
if necessary. Hence, the non - hermiticity does not prevent us from 
using the Dyson mapping directly. However, to avoid this technical
problem, in this paper we shall stick to the use of the quadrupole 
operator which generates hermitian images. 

For the sake of completeness, let's make some comments on the second
type of boson mappings mentioned above, the well  known Marumori 
method \cite{marumori}. It
focuses on the mapping of state vectors. This method defines
the operator in such a way that the matrix elements are conserved
by the mapping and the importance of the commutation rules is
left as a consequence of the requirement that matrix elements coincide
in both spaces.
The BZM and the Marumori expansions are equivalent 
at infinite order, which means that just with the proper mathematics
one can go from one expansion to the other.
 
{\it Quons} are particles whose statistics interpolates between the
boson and fermion statistics depending on a special deformation
parameter which varies from +1 to -1 \cite{green91}, \cite{green92}. 
This behaviour is easily
observed by the q- deformed commutation relations obeyed by the quons,
which appears in the next section. Notice that there are subtle
differences between the commutation relations of the usual quantum 
algebras and of the quons \cite{chaichian}. For q - bosons with zero
angular momentum the commutation relations are identical, but this
is not true for q-bosons with larger angular momentum.

In a recent work \cite{nosso}, we have shown that a deformed boson
mapping of the Marumori type, when applied to a pairing interaction,
provides excellent results, i.e., for a properly chosen deformation
parameter the exact result is also achieved.
 
In this work we concentrate on the Dyson mapping \cite{dyson}
for a nuclear single $j=5/2$ shell with identical particles. This
problem is different from others in the literature  \cite{mb}
in what concerns the boson commutation relations used during the mapping.
First of all, we briefly outline the main aspects of the mapping from
a fermionic space to a quantum deformed bosonic space. Once the 
deformation parameter is set equal to one, the usual boson expansion
is recovered.  Then
we give the expressions for the mapping 
of a quadrupole-quadrupole interaction, which is used for the 
comparison among the usual, the deformed mapping and the 
exact results obtained from a shell model calculation,
in the same spirit as in refs. \cite{my},\cite{menezes}. 

\section {The Generalized Dyson Boson Mapping}

The fermionic pair (${A^{\dagger}}^{J}_{M}$) and multipole ($B^{J}_{M}$) 
operators for a single-$j$ shell are defined as:

\be
 {A^{\dagger}}^{J}_{M}= \frac{1}{\sqrt2}
\sum_{m,m^{\prime}} (j~m~j~m^{\prime}|J~M)~
a^{\dagger}_{m} a^{\dagger}_{m^{\prime}} 
\ee
with $A^{J}_{M}=({A^{\dagger}}^{J}_{M})^{\dagger}$ and 

\be
B^{J}_{M} = 
\sum_{m,m^{\prime}} (j~m~j~-m^{\prime}|J~M)~
(-1)^{j -m^{\prime}}a^{\dagger}_{m} a_{m^{\prime}} ~~,
\ee

\noindent
where $a^{\dagger}_{m}~( a_{m^{\prime}})$ are fermionic  
creation~(annihilation) operators and  
the Clebsch-Gordan coefficients are defined as $(j~m~j~m'|J~M)$.
 It is straightforward to show 
that these operators satisfy the following 
commutation relations: 

\[
[B^{J_1}_{M_1},B^{J_2}_{M_2}]=\sum_J (-1)^{2j+J} 
[1-(-1)^{J_1 +J_2 +J}]\sqrt{(2J_1 +1)(2J_2 +1) }
\]
\be
\times ~(J_1 M_1 J_2 M_2 |J M_1 +M_2 ) 
 {{J_1~J_2~J}\brace j~j~j}
B^{J}_{M_1 +M_2 },
\label{cr1}
\ee

\be
[A^{\dagger~J_1}_{M_1},A^{\dagger~J_2}_{M_2}]=0,
\label{cr2}
\ee

$$
[A^{\dagger~J_1}_{M_1},A^{J_2}_{M_2}]= - \frac{1}{2} \delta_{J_1~J_2}
\delta_{M_1~M_2} (1-(-1)^{2j - J_2}) 
$$
$$
- \frac{1}{2} \sum_J
(-1)^{J+M_2} 
[(-1)^{J_1}-(-1)^{2j+J_2} - (-1)^{2j}- (-1)^{2j+J_1 +J_2}]
\sqrt{(2J_1 +1)(2J_2 +1) }
$$
\be
\times ~(J_1 M_1 J_2 -M_2 |J M_1 -M_2 ) 
 {{J_1~J_2~J}\brace j~j~j}
B^{J}_{M_1 -M_2 },
\label{cr3}
\ee

\[
[B^{\dagger~J_1}_{M_1},A^{\dagger~J_2}_{M_2}]= \sum_J
(-1)^{J+M_1} 
[(-1)^{2j} - (-1)^{J_2}] \sqrt{(2J_1 +1)(2J_2 +1) }
\]
\be
\times ~(J_1 -M_1 J_2 M_2 |J M_2 -M_1 ) 
 {{J_1~J_2~J}\brace j~j~j}
A^{\dagger~J}_{M_2 -M_1  }.
\label{cr4}
\ee

The usual Dyson mapping consists in the bosonic expansion of the 
pair and multipole operators, where the pair operators 
include only one and three-boson terms and the multipole operators 
only two-boson terms. The coefficients of the bosonic expansions are 
determined in order to preserve 
the same commutation relations for the mapped operators.  
 At this point we generalize the Dyson method, 
using q-deformed bosons in the Dyson 
bosonic expansion.   
The calculations in this work  will be restricted to a $j=5/2$ shell 
and will use only $s$,$d$ and $g$ bosons, which suffice to render the
mapping exact. 

Since we are interested in the quadrupole - quadrupole interaction,
we shall only explain explicitly how the q- deformed quadrupole
image is obtained. If we consider a single $j=5/2$ shell, the
commutation relations for this shell are found by substituting $j=5/2$ 
into eqs. (\ref{cr1}), (\ref{cr2}), (\ref{cr3}) and (\ref{cr4}).
The fermion multipole operators which appear in these commutation 
relations are going to be mapped onto functions of q- deformed
boson operators as follows:
$$B^J_M \rightarrow Q^J_M.$$
We want an exact mapping. Thus, we are going to use one q-deformed boson
for each possible pair operator and 
they satisfy the following  commutation relations\cite{green92}:
\be
[s,s^{\dagger}]_{q_{s}}=s~s^{\dagger}-q_{s}~s^{\dagger}~s=1,
\label{qs}
\ee
\be
[d_m,d^{\dagger}_{m^{\prime}}]_{q_d}=
d_m ~{\dd}_{m^{\prime}}-q_d ~{\dd}_{m^{\prime}} ~d_m = 
 ~{\delta}_{m,m^{\prime}},
\label{qd}
\ee
\be
[g_m,{\gd}_{m^{\prime}}]_{q_g}=
g_m ~ {\gd}_{m^{\prime}}- q_g ~ {\gd}_{m^{\prime}} ~ g_m =
~{\delta}_{m,m^{\prime}},
\label{qg}
\ee
In these equations the real numbers  $q_s$,$q_d$ and $q_g$ 
are defined as the deformation parameters associated 
respectively with the $s$, $d$ and $g$ 
deformed bosons. Note that 
we recover the usual bosonic (fermionic) algebra when the
three deformation parameters are equal 
to $ +1 (-1)$. The above commutation relations form an algebra 
 called {\it quon algebra} and it has been used to study  the
 small violations of Fermi or Bose statistics\cite{green91}.
As usual $s$,~$d$ and $g$ carry angular momentum 0,~2 and 4 
and $N_s,~N_d$ and $N_g $ 
are number operators associated with the $s$,~$d$ and $g$ q-bosons.

Now we present the Dyson expansion of the multipole operators of 
interest :
$$
Q^{2}_{M}=
A~[(\sd \otimes \dt )^{2}_{M} +(\dd \otimes s)^{2}_{M} ]
+B~(\dd\otimes \dt )^{2}_{M}$$ 
 \be{
+ C~[(\dd \otimes \gt )^{2}_{M} + (\gd \otimes \dt )^{2}_{M} ]
+~D~(\gd \otimes \gt )^{2}_{M}}
\label{Q2} 
\ee
\be{
Q^{1}_{M} = E~(\dd \otimes \dt )^{1}_{M}~+~
F(\gd \otimes \gt )^{1}_{M}
}\ee
\be{
Q^{3}_{M}=G(\dd \otimes \dt )^{3}_{M} +
H~(\gd \otimes \gt )^{3}_{M} +
~J~[(\dd \otimes \gt )^{3}_{M} + (\gd \otimes \dt )^{3}_{M} ]
}\ee
where we have used the notation for the spherical tensors :
$${\dt}_{-m}=(-1)^{m}~d_m~~~~,~~~~{\gt}_{-m}=(-1)^{-m}g_m $$
 and 
\[
[T^{j_1} \otimes T^{j_2}]^{J}_{M} = 
\sum_{m,{m^\prime} }  (j_1~m~j_2~-m^{\prime}|J~M)~T^{j_1}_{m}~
T^{j_2}_{m^\prime}.
\]

The coefficients $A,B,C,D,E,F,G,H,J$ are determined imposing 
that $Q^{J}_{M}$ satisfy the commutation relations obtained from the
fermionic operators.
Next we outline the steps followed in obtaining the coefficients
of the multipole operators, bearing in mind that all commutation
realtions used here come from eq. (\ref{cr1}). 
From $[B^1_{M},B^1_{N}]$ we have obtained $E^2$ and $F^2$,
from $[B^2_{M},B^2_{N}]$ we have obtained $A^2$,$B^2$, $C^2$,
$D^2$, $BC$ and $CD$ and
from $[B^3_{M},B^3_{N}]$ we have gotten $GJ$, $HJ$, $J^2$,
$H^2$ and $G^2$.
Of course many other commutation relations can be used to
check the results of the coefficients once the number of 
commutation relations is larger than the amount of coefficients
to be determined. We have finally obtained:

\begin{tabular}{lll}

A= 0.8165   & B=-0.5832  & C= 0.9091 \\
D= 0.8207   & E= 0.7559  & F= 1.8516 \\
G=-0.9091   & H=-0.3350  & J=-1.235 .   

\end{tabular}

Notice that these coefficients do not depend on $q$, although
expressions (\ref{qs}),(\ref{qd}) and (\ref{qg}) have been
used together with the Wick contraction technique for their
calculation. They also agree with the results coming from
the general formula for the usual Dyson mapping coefficients \cite{zb},
\cite{mb}. Nonetheless, the quadrupole- quadrupole interaction
does depend on $q_s$, $q_d$ or $q_g$, as can be seen below. 

From the expression  of $Q^{2}_{M}$ given in eq.(\ref{Q2}) 
it follows that the Dyson mapped quadrupole-quadrupole operator
is given by: 

\[
Q.Q = \sum_{M} (-1)^{M}~Q^{2}_{M}~Q^{2}_{-M} =
\]
\[
A^2 \sd ~s + (A^2  +B^2  +C^2)
\dd \cdot \dt 
\]
\[
+{5\over 9}(C^2~+~D^2~) \gd \cdot \gt 
\]
\[
+(A^2 
~\sd ~\sd \cdot~\dt \dt ~+~ h.~ c.)
+A^2(q_d + q_s)~\sd \dd \cdot \dt s  
\]
\[
+5~ B^2 q_{d}~\sum_{J}  {{2~2~2}\brace 2~2~J}~
~[\dd \otimes \dd ]^{J} \cdot [\dt \otimes \dt ]^{J}  
\]
\[
+ (5C^2 \sum_{J} {{4~2~2}\brace 2~4~J}~
 [\gd \otimes \gd ]^{J} \cdot  [\dt \otimes \dt ]^{J}  +~h.~c.)
\]
\[
+5~(~(q_d +q_g )C^2  \sum_{J} {{2~4~2}\brace 2~4~J}~ +
2BD \sum_{J} 
{{2~2~2}\brace 4~4~J}~)~
[\dd \otimes \gd ]^{J} \cdot [\dt \otimes \gt ]^{J}
\]
\[
+5~ D^2 q_{g}~\sum_{J}  {{4~4~2}\brace 4~4~J}~
~[\gd \otimes \gd ]^{J} \cdot [\gt \otimes \gt ]^{J}  
\]
\[
+(AB~(q_d+1)\sd \dd \cdot [\dt \otimes \dt ]^{2} ~+~h.~c.)
\]
\[
+(AC~(q_{d} + 1 ) \sd \dd \cdot [\dt \otimes \gt]^{2} 
~+~h.~c.)
\]
\[
+ (AC~{{2 \sqrt{5}}\over{3}}~\sd \gd \cdot [\dt \otimes \dt ]^{4} 
~+~h.~c. ) 
\]
\[
+(BC~5~(1+ (-1)^J~q_{d})\sum_{J} {{4~2~2}\brace 2~2~J}~ [\dd \otimes \gd ]^{J} 
\cdot  [\dt \otimes \dt ]^{J}  ~+~h.~c.) 
\]
\[
(CD~5~(1+ (-1)^J~q_{g})\sum_{J} {{4~2~2}\brace 4~4~J}~ [\gd \otimes \dd]^{J} 
\cdot  [\gt \otimes \gt ]^{J}  ~+~h.~c.) 
\]
\be
+(AD~{{2\sqrt{5}}\over{3}}~\sd 
\gd \cdot [\dt \otimes \gt ]^{4} 
+~h.~c. )
\ee

where $Q^{2}_{M}$ is defined in eq.(\ref{Q2}) and we have used the 
notation:
 
$$T^{J}~\cdot ~U^{J}=(-1)^{J}\sqrt{2J+1}~[T^{J}\otimes U^{J}].$$

Thus, we have obtained the q-deformed Dyson mapping of the 
quadrupole - quadrupole operator.

\section{Applications and Results}

In order to study the properties of the q-deformed Dyson expansion 
we propose to compare the diagonalization of a quadrupole 
hamiltonian mapped onto the bosonic subspace generated by 
$s,d$ and $g$ 
bosons with the exact 
shell model calculation restricted to four fermions in a $j=5/2$ shell. To achieve our goal, one has to introduce a basis for the diagonalization of the desired hamiltonian, which consists of two q- deformed boson states given by

\be
|ss>= \frac{1}{\sqrt{(1+q_s)}} s^{\dagger}~ s^{\dagger}~ |0>
\ee
\be
|dd>=\frac{1}{\sqrt{(1+(-1)^{\lambda}~q_d)}} \sum_{m,m'} (2~ m~ 2~ m'~|\lambda~ m+m')~
d^{\dagger}_m~ d^{\dagger}_{m'}~ |0>
\ee
\be
|gg>=\frac{1}{\sqrt{(1+(-1)^{\lambda}~q_g)}} \sum_{m,m'} (4~ m~ 4~ m'~|\lambda~ m+m')~
g^{\dagger}_m~ g^{\dagger}_{m'}~ |0>
\ee
\be
|sd>= s^{\dagger}~ d^{\dagger}_{m}~ |0>
\ee
\be
|sg>= s^{\dagger}~ g^{\dagger}_{m}~ |0>
\ee
\be
|dg>=\sum_{m,m'} (2~ m~ 4~ m'~|\lambda~ m+m')
d^{\dagger}_m~ g^{\dagger}_{m'}~ |0>
\ee

\noindent where $\lambda$ is the total angular momentum of the basis state. 

     The main results of our diagonalization can be summarized
 in Fig. I. There we make an analysis of the behavior of  the q- deformed bosonic spectrum compared with the one obtained through the exact shell-model diagonalization procedure for four fermions in a single $j=5/2$ shell. In this simple case, we get just three states for the fermionic spectrum with angular momenta 0,2 and 4. The bosonic results however are highly dependent on the values of the q- parameters. We start by comparing the spectrum obtained making $q_s=1, q_d=1$ and $q_g=1$, i.e., turning off the effects of the deformation. In this case the agreement is very poor. Also, note that the spurious part of the bosonic spectrum lies well above  the "physical" states.

     Instead of doing just a fit to the exact spectrum in order to find out the best values for the q parameters, we have decided first to try to find a criterion for fixing those parameters. In this work we are considering a relatively low value for the shell angular momenta $j$, which is known not to be a very favorable situation for the quadrupole- quadrupole
boson expansion. However, it is our intention to test our deformed bosonic expansion even for that not so favorable case and show that our description can improve considerably the results. We start by considering again the pair operator $A^{\dagger~J}_{M}$ defined in equation (1) and identifying it with the boson deformed operators $s, d$ and $g$ respectively for $J=0,2$ and 4 in analogy with reference \cite{russians}. Of course, the pair operator cannot be treated as a real boson operator and this can be made clear, for example, when we take the norm of the state:

$$ \frac{1}{\sqrt{2}}[A^{\dagger~J} \otimes A^{\dagger~J}]^{\lambda}_{\mu}|0>=\frac{1}
{\sqrt{2}}\sum_{m,m'}(J m J m'|\lambda \mu) A^{\dagger~J}_{m}A^{\dagger~J}_{m'}|0>. $$

    Note that the state considered in the above equation has, according to our assumption, the bosonic images defined in equations (14), (15) and (16). Those are the basis states that define the ground state of the system taking $\lambda =0$. On the other hand one may easily show that:
\be
{\frac{1}{2}<0|[A^{J} \otimes A^{J}]
^{0}_{0}[A^{\dagger~J} \otimes A^{\dagger~J}]^{0}_{0}|0> = 1-2(2J+1)^{2}\left \{\begin{array}{clcr} j&j& J\\ j&j&J \\ j&j&0 \end{array} \right \}}.\ee

      The term depending on the 9-J simbol expresses the effect of the Pauli principle and its magnitude might be used to estimate the validity of our bosonic identification between the pair and deformed boson operators. Considering now the norm of the two-boson states:

\be {\frac{1}{2}<0|sss^{\dagger}s^{\dagger}|0>=1+\frac{q_{s}-1}{2}}
\ee

\be {\frac{1}{2}<0|[d \otimes d]^{0}_{0}[d^{\dagger}\otimes d^{\dagger}]^{0}_{0}|0>=1+\frac{q_{d}-1}{2}}\ee

\be {\frac{1}{2}<0|[g \otimes g]^{0}_{0}[g^{\dagger}\otimes g^{\dagger}]^{0}_{0}|0>=1+\frac{q_{g}-1}{2}}\ee

\noindent and comparing the three above equations with equation (20) we may get an estimation for the parameters $q_{s},q_{d}$ and $q_{g}$. For $j=5/2$ we obtain $q_{s}=0.3$ and $q_{d}=0.6$. For $q_{g}$ however we do not find a reasonable value (between -1 and 1). This seems to be a problem inherent to a small j-shell value, once we check that for bigger values of $j$ we allways end up with reasonable values for the three q parameters (including j=7/2). So we decide to keep $q_{g}=1$. Observing again Fig. I, one can see that our choice for the parameters improves considerably the agreement with the exact shell model calculation.   

 Finally, turning back to the simple fitting procedure, we search for the best matching between fermionic and bosonic levels varying two deformation parameters, while keeping the third one fixed . We find that the best agreement is reached when we fix the g-boson deformation parameter ($q_g=1$), assigning to $q_s$ and $q_d$ the values 0.2 and 0.4 respectively, as also shown in Fig. I. Of course we also might
vary independently the three parameters. However, we have checked numerically that the quality of the fitting is not considerably improved in that case.  

    In summary, in this work we find the deformed bosonic image of the quadrupole-quadrupole hamiltonian through the implementation of a q - deformed Dyson mapping based in the so-called {\it quon} algebra. Explicit results are presented for a system of four identical fermions in a single $j=5/2$ shell and the bosonic expansion is written in terms of deformed $s, d$ and $g$ bosons. We then try to find a way to fix the q parameters associated with each boson operator by their identification with the pair operator carrying the corresponding angular momenta. Compared with the usual non-deformed result for the energy spectrum, a remarkable improvement is achieved using this method. However, for the $g$ boson we do not find a reasonable value for the deformation parameter in the case considered here. As for larger shells and the same number of fermions we can find reasonable q values for the three parameters using that very same method, we believe that it is worthwhile to proceed!
  further this investigation in a 

\vskip 0.35in

     This work has been partially supported by CNPq.

\newpage
\section{ Figure Captions}

  FIGURE I.  Results for the energy levels for a system composed of four particles in a j=5/2 shell interacting through the quadrupole-quadrupole force in a) the exact shell-model diagonalization, b)using the deformed Dyson boson expansion of this work with $q_{s}=1.0, q_{d}=1.0$ and $q_{g}=1.0$,
c)using $q_{s}=0.3, q_{d}=0.6$ and $q_{g}=1.0$ (physical interpretation) and d)$q_{s}=0.2, q_{d}=0.4$ and $q_{g}=1.0$ (best fitting). See the text for details.


\begin{thebibliography}{99}
  
\bibitem{bzm} S.T. Beliaev and V.G. Zelevinsky, Nucl. Phys. 39
(1962) 582 ; E.R. Marshalek, Nucl. Phys. A161 (1971) 401,
A224 (1974) 221, 245.
 
\bibitem{dyson} F.J. Dyson, Phys. Rev. 102 (1956) 1217;

\bibitem{marumori} T. Marumori, M. Yamamura and A. Tokunaga,
Progr. Theor. Phys. 31 (1964) 1009; T. Marumori, M. Yamamura,
A. Tokunaga and T. Takada, Progr. Theor. Phys. 32 (1964) 726.

\bibitem{green91} O.W. Greenberg, Physical Review D 43 (1991) 4111.

\bibitem{green92} O.W. Greenberg, Physica A 180 (1992) 419.

\bibitem{chaichian} M. Chaichian, R. Gonzalez Felipe and C. Montonen,
J. Phys. A 26 (1993) 4017

\bibitem{nosso} S.S. Avancini, F.F. de Souza Cruz, J.R. Marinelli,
D.P. Menezes and M.M. Watanabe de Moraes, J. Phys. A 29 (1996) 5559

\bibitem{mb} D.P. Menezes and D. Bonatsos, Nucl. Phys. A 499 (1989) 29.

\bibitem{my} D.P. Menezes and N. Yoshinaga, Phys. Lett. B 221 (1989) 103.

\bibitem{menezes} D.P. Menezes, Phys. Scripta 45 (1992) 302.

\bibitem{zb} M.R. Zirnbauer and D.M. Brink, Nucl. Phys. A 384 (1982) 1;
 M.R. Zirnbauer, Nucl. Phys. A 419 (1984) 241.

\bibitem{russians} R. J. Jolos, I. Kh. Lemberg and V. M. Mikhailov, Fiz. Elem. Chastits At. Yadra 16 (1985) 280.

\end{thebibliography}
\end{document}